# On the theory of moveable objects

*Abstract*. User-driven applications belong to the new type of programs, in which users get the full control of WHAT, WHEN, and HOW must appear on the screen. Such programs can exist only if the screen view is organized not according with the predetermined scenario, written by the developers, but if any screen object can be moved, resized, and reconfigured by any user at any moment. This article describes the algorithm, by which an object of an arbitrary shape can be turned into moveable and resizable. It also explains some rules of such design and the technique, which can be useful in many cases. Both the individual movements of objects and their synchronous movements are analysed. After discussing the individually moveable controls, different types of groups are analysed and the arbitrary grouping of controls is considered.

## Content

## Introduction

Moving of the screen objects is an extremely important thing, so the efforts in this direction were already made years ago. Any reader of this article can remember the programs, in which he saw the moveable objects. But the expression *moving of the screen objects* can mean absolutely different things, so let's make the purpose of my research and the results, which are described here, absolutely clear.

When a screen object is moved regardless of the users' actions or as the result of his pressing the keyboard, but the reaction on each click is totally predetermined at the stage of the program development, then it is an animation. The moving, which is absolutely predicted and fixed in the code of a program by its developer, is not the purpose of this work.

There are also movements, which are not predefined, but are not the real movements, though they can look like movements as a result of simple programmers' tricks (or knowledge, if you look at it from slightly different angle). For example, in several programs (the well known *Paint* is one of them) you can select a rectangular area by pressing the mouse at one point and dragging the mouse across the screen without releasing; while the mouse is moving, the dashed line moves across the screen. Even with the much slower computers, than we have now, this trick was easy to implement by using the XOR operation for drawing; it is not the moving of an object.

By moving of the screen objects I mean the process, when user selects an object and moves it across the screen according with his own wish. The main difference from the previously mentioned cases is that such moving can't be predicted by the developer, because it is decided by the user only at the time, when the application is running. The real moving of some objects was introduced in some of the older programs, but in each case it was especially done for one or another particular class of objects. My opinion that for a specific class of objects everything can be done; it's only the question of time and efforts (brains are also among the needed ingredients). But the existence of those rare samples of really moveable objects underlines even stronger a couple of very important things.

1. We really need the ability to move the screen objects, because it is the most natural thing, which we do with all the objects around us in our normal (not virtual) life.

2. The problem of moving the screen objects is not trivial; that's why it was solved so rarely and only for a limited set of objects.

Even the programs, which introduced such classes of moveable objects, did it for some special tasks. There was not a single program, in which <u>everything was moveable and resizable</u>. I work on development of very sophisticated applications for many years and I never heard about a single application, in which everything would be moveable. I especially underlined not only the important, but the mandatory (absolutely required) feature of the user-driven applications.

This and other features of such applications will be the subject of the second article (strongly related to this one), but the crucial point in development of such applications is the algorithm that allows to turn any screen object into moveable / resizable.

What are the basic requirements to such algorithm?

**First**, it must be easy. I don't know anything about the complexity of the algorithms, used by other people before, but I know that when I did similar things (moving some objects around the screen) years ago, those algorithms were not simple at all. Those algorithms were understandable by the author (me!), but I wasn't even thinking about spreading them around. I assume that at the same level of complexity were the algorithms, on which other samples were based. The best procedure of turning objects into moveable was once described by Charles Perrault[*]; I hope to design an algorithm, which is not far behind that one. With that famous example, we know that it looked simple, when it was already in use, but there is no information, on how many years of preliminary work it required. The same thing with my algorithm: you'll see further on that it is extremely easy in use, but it took years of thinking and work to design it to such level.

**Second**, the algorithm must be applicable to any shape of objects. One of the things that people immediately begin to think about, when they need to move some objects across the screen, is the use of some bitmap operations. Those operations are applicable to the rectangular areas, and it's not an algorithm, but a trick with serious limitations. The good algorithm must work with an arbitrary shape of objects. While showing the samples (there will be a lot of them) further on in the article, I'll demonstrate some cases, which are simple in design; others will need a bit of extra thoughts, but there are no restrictions on the form of the moveable objects. An object can even consist of the separated parts; still the algorithm works without any problems.

**Third**, the algorithm doesn't require any visual changes to the objects. The screen objects are designed according with the ideas of their developers; the algorithm must not interfere with this process. It must be enough for users to know that the objects are moveable / resizable. Though, if the designers want, they can add some indication of these features. The algorithm must work regardless of these visual indications.

**Fourth**, these features (moveable and resizable) are not applied to the objects by the designer only. They can be changed by the users according with the purpose of application. For example, if the application is used for assembling the complex object of some parts, then the user can easily move and change all these parts, but when he decides that the result is satisfactory, then he can declare this object an unchangeable (but, for example, still moveable) in order to prevent any accidental changes.

**Fifth**, there are no restrictions in the way the algorithm must be used. There are some of my suggestions; I tried to make all of them reasonable, but none of them is a mandatory requirement.

Here are these suggestions.

- From my point of view the most natural way of moving / resizing the screen objects is to use the mouse. It's the most accurate way of the object's selection, placing, and reconfiguring regardless of its shape, size or position.

- To move an object, you simply press it at any point and drag to another place on the screen.

- To resize an object, you press it on the border and either resize it, or reconfigure. It's a challenge for the designer to make the resizing absolutely natural and according with the users' expectations. But this is one of the things, in which the designer's experience and understanding of the situation means a lot. In reality this is a difference between the good design and the bad one. Though the resizing by the border is absolutely natural, there can be nuances even in the simple cases. For example, the resizing of a circle means the symmetrical change in all the directions, but what about a rectangle? Is there going to be a symmetrical change on the opposite side, or the rectangle must be resized only in the direction of the moved border? This can depend on the application and on the objects, but this sample shows that even the simplest cases can be not so obvious, as it looks.

---

[*] C. Perrault, Cendrillon, 1697

- In all my applications, the left button is used to start the forward movement or resizing, the right button is used to start the rotation. This is not fixed in any of the proposed classes or anywhere else. It's for anyone to decide, though whenever I see the implementation of the forward movement (drag-and-drop) in any other application, it is always done by the left button. Windows on the upper level is the best and most obvious example of using the left mouse for moving and resizing. If the left button is used for forward movements, resizing, and reconfiguring, then the use of the right button for rotation is the most natural thing. (There is only one class in my samples – a class of rotatable texts, for which the rotation with the right button was put into code. This was done only to exclude the duplication of the same several lines of code in all the places, where I use this class. Anyway, the class is primitive and can be easily reproduced without such limitation.)

- The exclusive use of the mouse without any additions is enough for all the purposes even in the most complex classes. Because of this, all my samples now use only `MouseDown`, `MouseMove`, and `MouseUp` events. It's again not the requirement that is strictly coded somewhere inside, but only my suggestions.

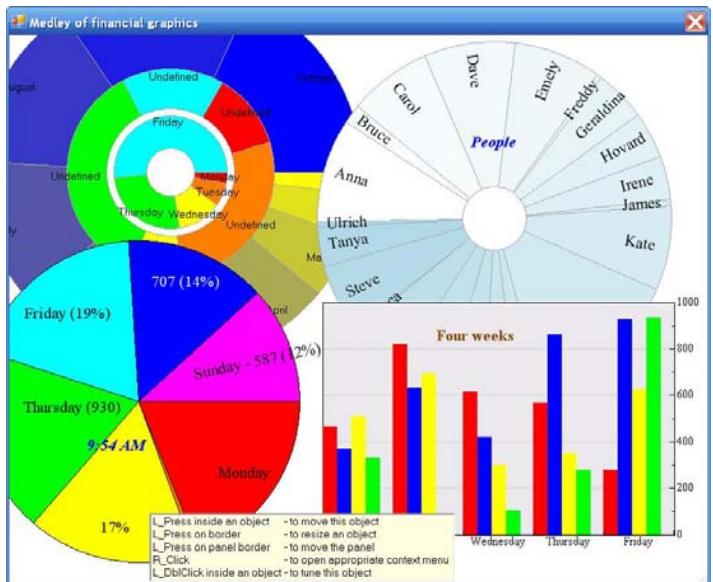

**Fig.1** Different types of moveable / resizable plotting

The samples, which I'll use for explanation, are made as simple as possible. A majority of the samples for this article will be from the set of simple geometrical figures, but this is done only for the purpose of better explanation. Don't think that the use of the proposed algorithm is limited by this set of figures. The inspiration for this work came from the huge demand for such moveable / resizable graphics for the area of the most complicated scientific applications, and this is where the results are used now all the time. The more complicated is the task, the more extraordinary are the results of using this algorithm. It can be the area of the engineering / scientific applications, or it can be for some financial analysis. **Figure 1** shows the view from one of such demo applications.

In this form, any number of plots of different types can be added to the screen by the user at any moment. All of them can be moved, resized or rotated. Each piece of textual information can be moved and rotated individually. All the visualization parameters of the objects can be tuned individually or spread on the group of "siblings". I don't want to go into all the details and features of such applications just now; I'll write about it in the second article, but the main point is that exactly the same algorithm, which will be described on the simple samples, is applicable to the most complicated classes of objects. And as you'll see, even the complexity of using the algorithm doesn't go up in stepping from the primitive figures into the world of the scientific plotting. I would say more: the simplicity is exactly the same.

Throughout this article, I'll be using the samples only from the **TheoryOfMoveableObjects** program (**figure 1** is an exception). This application was especially designed to accompany this article. The application, in the form of a ZIP file, is available at www.SourceForge.net in the **MoveableGraphics** project (names of the projects are case sensitive there!). The ZIP file contains the whole application's project, written in C# and ready to be used in Visual Studio. If you want only to run this application to see all the things, I am going to write about, then you need only two files from that ZIP file: **TheoryOfMoveableObjects.exe** and **MoveGraphLibrary.dll**.

A short, but important remark about this program: there is not a single unmoveable object in this application.

I constantly work with the C# language, so all the programs, I am going to mention here, are written in this language; the code samples will be in C#, and I'll use some terms from this language. But the algorithm and designed classes are not linked to C# – they can be easily developed with other instruments; I simply prefer to use C#.

This paper consists of several parts.

Part 1  introduces the terms and the algorithm.

Part 2  describes the moving and resizing of the simple figures, introduces the technique of resizing the objects with the straight and curved borders, and looks into the world of rotation.

| Part 3 | explores more complicated objects, in which the parts can move individually, but all of them can be involved in synchronous movements. |
|---|---|
| Part 4 | deals with the controls and groups of controls, which are used as the blocks in forms' design; also demonstrates the sample of an arbitrary controls' grouping, which is decided by the users. |
| Conclusion | contains the summary, but also works as a link to the second article. |

# Algorithm

My idea of making graphical objects moveable and resizable is based on covering an object with a set of sensitive areas, which are used either to start the moving of the whole object or some part of it. These areas are called ***nodes*** and belong to the `CoverNode` class. The nodes are usually invisible, but their movements are transferred into real movement of the screen object, with which they are associated. Four different types of movement are considered.

- If the size of an object is not changed and an object is moved without any change of relative positions of its parts and without rotation, then it is a forward movement.
- If all the parts of an object are increased or decreased with the same coefficient, but the general shape is not changed, then it is a resizing.
- If a movement of any part changes the relative positions of the parts and the general shape, then it is reconfiguring.
- If the general shape is not changed, but the whole object is turned from the original position, then it is a rotation.

The first three movements (forward movement of the whole object, resizing or reconfiguring) are started with the left button press; the choice between these movements is decided by the starting point and the node, to which this point belongs. There are special areas to start reconfiguring (it depends on the object) and resizing (usually it is the border); at any inner point of an object the forward movement of the whole object is usually started. The start of rotation is distinguished from other three movements by using the different button (the right one). The rotation of an object usually starts at any point.

There are no limits on the size of the nodes, but their possible shapes are described by this enumeration.

```
enum NodeShape { Circle, Polygon, Strip };
```

Any circular node is defined by its central point and radius.

Polygon node is described by an array of points, representing its vertices. Polygon must be convex!

A strip node is defined by two points and radius. The strip node can be also looked at as a rectangle with two additional semicircles on the opposite sides. Those two points are the middles of those sides and the centers of the semicircles; the diameter of the semicircles is equal to the width of the strip.

In addition to the location, size, and shape, each node has a parameter, which determines the shape of the mouse cursor, when the mouse is moved across this node. Usually this is the only prompt for the users that they can grab an object under the mouse and move it one way or another, so it's better to make the cursor's shape informative and not confusing. Throughout all the further samples you'll find out that if the rectangular object can be resized in the horizontal or vertical direction, then the `Cursors.SizeWE` and `Cursors.SizeNS` are used; the resizing possibilities for objects of an arbitrary form are signaled by the `Cursors.Hand`; the possible movement of the whole object is signaled by the `Cursors.SizeAll`.

Any node has one more parameter; originally it was used for declaring the possibility of the node's individual movement. In reality these movements are defined by the code of the `MoveNode()` method, which is described a bit later. But this parameter turned out to be of an extreme value in cases of the objects with nontrivial shape and in situations, when user wants to change the object's behaviour. Some of these possibilities will be described further on in this article; others will be mentioned in the second article, while talking about very complicated objects.

An array of nodes, covering an object, is called a ***cover*** (class `Cover`). A cover must include at least one node. There are no other restrictions on the number of nodes, their types or sizes. The relative positions of the nodes are not regulated at all. If you want an object to be moveable by any point, then the whole object's area must be covered with the nodes; any gaps in such situation will result in the appearance of the places, by which an object can't be moved. The overlapping of the nodes is not a problem at all, but their order can be important, if they are used for different movements, as the decision is made by analyzing the nodes according with their order in the cover. Usually the number of nodes is small even for very complicated objects, but there are cases, when a significant number of nodes is needed. This often happens for the objects with the curved borders; covers of such type are called the ***N-node covers***.

Technically I saw two ways to add moveable / resizable features to the objects: either to use an interface or an abstract class; after trying both ways I decided upon an abstract class. Any graphical object that you want to turn into moveable and resizable must be derived from the abstract class `GraphicalObject`, and three crucial methods must be overridden

```
public abstract class GraphicalObject
{
    public abstract void DefineCover ();
    public abstract void Move (int dx, int dy);
    public abstract bool MoveNode (int i, int dx, int dy, Point ptMouse,
                                   MouseButtons catcher);
```

For graphical objects, the cover is organized in the **DefineCover**() method. Controls are wrapped by a graphical object of the `FramedControl` class; for controls none of these methods is needed, as everything is automated.

Each node in the cover has its personal identification number (from the range between 0 and the number of nodes in the cover); this number is often used for the node identification, while the decision about movement is made in the `MoveNode()` method.

**Note**. Though the whole idea of moving and resizing objects is based on cover, the `DefineCover()` method is usually the only place, where you have to think about the cover! The rule is: organize the cover and forget about it. There is one exception of this rule; which is mentioned several lines below in the explanation of the `MoveNode()` method.

**Move** (dx, dy) is the method for <u>forward moving of the whole object</u> for a number of pixels, passed as the parameters.

The drawing of any graphical object with any level of complexity is usually based on one or few very simple elements (`Point` and `Rectangle`) and some additional parameters (sizes). While moving the whole object, the sizes are not changed, so only the positions of these basic elements have to be changed in this method.

**MoveNode** (i, dx, dy, ptMouse, catcher) is the method for <u>individual moving of the nodes</u>. The method returns a Boolean value, indicating whether the required movement is allowed; in the case of forward movement, the `true` value must be returned if any of the proposed movements along the X or Y axes is allowed. If the movement of one node results in relocation of other nodes, it is natural to call the `DefineCover()` from inside this method, and then it doesn't matter, what value is returned from the `MoveNode()`. The call for the `DefineCover()` from inside the `MoveNode()` method may happen even for the forward movement, when movement of one node affects the relocation of other nodes, and it usually happens with rotation, when all the nodes must be relocated. This is the exception of the previously mentioned rule, that cover is not even thought about anywhere outside the method of its definition.

Though the call of the `DefineCover()`method from inside the `MoveNode()` method looks like a very good idea, it has some limitations of its own. In some cases, the movement of a node starts the resizing of an object, which in turn demands the change of the number of nodes in the cover (this depends on the `Cover` design); in such situation, the call to the `DefineCover()` method from inside the `MoveNode()` method may cause a problem, because in the new cover the node with the same number (the one, which was originally pressed and caught) may belong to the node with different rules for moving. To avoid these problems, in such cases the cover is not changed throughout the resizing, but only when an object is released; the explanation of such a case with more details is in the chapter *N-node covers*.

`MoveNode()` method has five parameters:

| | |
|---|---|
| `int i` | Identification number of the node – the same number that was used in the `DefineCover()` method on design of this node. |
| `int dx` | Movement (in pixels) along the horizontal scale; positive number for moving from left to right; use this parameter if you write the code for forward movement. |
| `int dy` | Movement (in pixels) along the vertical scale; positive number for moving from top to bottom; use this parameter if you write the code for forward movement. |
| `Point ptMouse` | The mouse cursor position. For calculations of forward movement I simply ignore this parameter and use the previous pair; for calculations of rotation I ignore the previous pair and use only this mouse position; I found it much more reliable for organizing any rotations. |
| `MouseButtons catcher` | Informs you, which mouse button was used to grab the object. If by the logic of an application the object can be grabbed by any mouse button, then ignore this parameter; if the move is allowed by one button only, then use this parameter; in case the node can be involved in both types of movement (forward movement and rotation), this is a very useful parameter to distinguish between them. |

The `MoveNode()` method can be the longest of all three, because it must include the code for moving each node, but it is always uncomplicated, as more often than not the code for different nodes is partly the same. There are interesting situations, when the `MoveNode()` method is very short, though the number of nodes is big. For example, the *N-node* covers often consist of a significant number of nodes. For such covers, the `MoveNode()` method is usually very short, because the behaviour of all those nodes is identical.

Even if an object is derived from the `GraphicalObject` class, thus receiving an ability to become moveable / resizable, it can be really moved and resized only if it is registered with the ***mover*** (of the `Mover` class). Mover supervises the whole moving / resizing process and in addition can provide a lot of information, associated with it. Regardless of the number of different objects, involved in moving / resizing, usually there is a single mover per form (dialog). However, there are situations when it is better (easier and more reliable) to have several movers in the form; for example, when you have moveable objects on different panels.

Only three mouse events - `MouseDown`, `MouseMove`, and `MouseUp` - are used for the whole moving / resizing process, and these are the methods, where mover works. To organize the moving / resizing of the objects in the form, several steps must be made.

Declare and initialize a `Mover` object:

```
Mover mover;
mover = new Mover (this);
```

Register with the mover all the objects that you want to move and / or resize

```
mover .Add (…);
mover .Insert (…);
```

Add the code for three mouse events

```
private void OnMouseDown (object sender, MouseEventArgs e)
{
    mover .Catch (e .Location, e .Button);
}
private void OnMouseUp (object sender, MouseEventArgs e)
{
    mover .Release ();
}
private void OnMouseMove (object sender, MouseEventArgs e)
{
    if (mover .Move (e .Location))
    {
        Invalidate ();
    }
}
```

This is not a simplification; this is the code, which you can see in nearly every form of the demo applications, regardless of the number or complexity of the objects, involved in moving / resizing. The three calls to three mover methods (one call per each mouse event) are the only needed lines of code! Further on you'll see some additional lines of code inside these methods, but they are used only for some auxiliary things, like changing the order of objects on the screen or calling different context menus on different objects.

Those three mouse events – `MouseDown`, `MouseMove`, and `MouseUp` – are the standard and often the only places, where mover is used. There are two other events – `MouseDoubleClick` and `Paint`, where mover can be mentioned and used, but this happens only on special occasions.

I often use the `MouseDoubleClick` event for opening the tuning forms of the complex objects, for example, scales and different plotting areas (as seen at **figure 1**). Without mover at hand (before implementing the moving / resizing of objects), I had to decide about the clicked object by comparison of the mouse location and the boundaries of the objects. Mover can do this job much better, as it informs not only about the occurrence of any catch, but also about the class of object, which was caught. And as any object, derived from the `GraphicalObject`, gets a unique identification number, then with this I get an easy access to the object itself.

```
private void OnMouseDoubleClick (object sender, MouseEventArgs e)
{
    if (mover .Catch (e .Location, MouseButtons .Left)) {
```

```
          … …
       if (mover [iInMover] .Source is MSPlot)
       {
```

Inside the `Paint` event, mover can be mentioned on those rare occasions, when covers have to be visualized, but it's really a rare thing, as good samples of design are those, which do not require such visualization.

```
    private void OnPaint (object sender, PaintEventArgs e)
    {
        mover .DrawCovers (e .Graphics);
```

## Safe and unsafe moving and resizing

The design of applications on the basis of moveable elements opens an opportunity for accidental disappearance of the elements from view.  This never happens in the ordinary applications with the unmoveable elements, but now users can simply move any element out of view across the border, release it there, and then what?  If in the resizable form an object is moved across the right or bottom border, then it's not a problem, as the form can be enlarged and the object returned back into play.  Such temporary relocation of the currently unneeded objects is often used in the user-driven applications.  But if an object is moved anywhere across the upper or left border of the form and dropped there, then there is no way to return it back into view by resizing the form.  The mover can take care of this situation and prevent such disappearance of the objects, but only if you want it to overlook this process.  For this purpose, the mover has to be initialized with an additional parameter – the form itself.  (If the mover works on a panel, then this panel is used as a parameter.)

```
    mover = new Mover (this);
```

You can find throughout the code of the **TheoryOfMoveableObjects** application that such type of mover's initialization is used in all the forms.  In such a case, when mover grabs any element for moving or resizing, the clipping will be organized inside the borders of the client area.  The level of clipping can be changed with one of the `Mover`'s properties; three levels of clipping are implemented

```
        public enum Clipping { Visual, Safe, Unsafe };
```

- `Visual` – elements can be moved only inside the client area.
- `Safe`   – elements can be moved from view only across the right and bottom borders of the form.
- `Unsafe` – elements can be moved from view across any border.

By default, the `Visual` level is used, but can be changed at any moment.  No clipping is used, if the mover is initialized without that additional parameter.

The unlimited shrinking of an object can lead to its total disappearance.  To avoid this, the minimum sizes must be declared for any class of resizable objects; these restrictions are used in the `MoveNode()` method.

**Technical note.**  To avoid the screen flicker, don't forget to switch ON the double-buffering in any form, where you use the moving / resizing algorithm.  It has nothing to do with the described technique, but it is a nice feature from Visual Studio.

## The simplest case – single node covers

Covers may consist of an arbitrary number of nodes.  The minimum number of nodes in the cover is one, so let's look at the simplest case of objects, which cover consists of a single node. As there are three possibilities of the node shapes, we'll have a representative of each type (**figure 2**).  In addition to three colored figures, the **Form_Nodes.cs** (menu position *Nodes and covers –*

*Node shapes*) includes two more objects.  The button  can be moved by its border; I'll write about the moveability of the individual controls and the groups of controls further on in the special chapter.  By clicking this button, it is possible to switch ON / OFF the visualization of covers.  The text, which you can see in this form, belongs to the `TextM` class; such objects can be moved by any inner point.  The cover of such objects also consists of a single node.  All the informative texts in this application are organized in such a way.

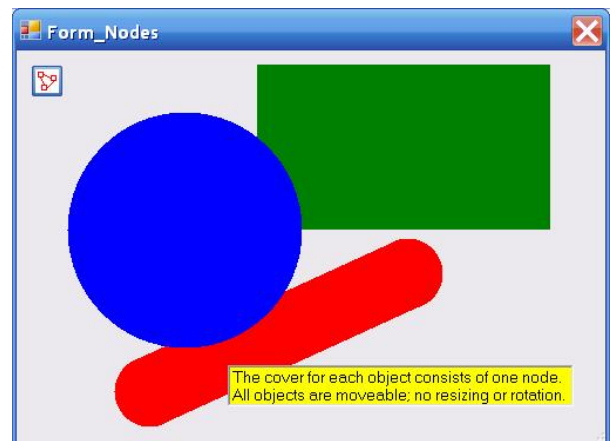

**Fig.2**  Three shapes of nodes

Each node of any cover can be used either to start the moving of the whole object or the resizing (reconfiguring). When a cover consists of a set of nodes (all further samples in this article will be of such a type), then some of the nodes are used for resizing, others – for moving. Because the cover of each colored figure in this form consists of exactly one node, and these objects are moveable, then those single nodes are used to start the moving. Thus all three objects are not resizable. Further on I'll demonstrate the resizable figures of the same shapes, but each of their covers will include a set of nodes. As the colored figures in the **Form_Nodes.cs** can be only moved and nothing else; I included into the names of their classes the word *primitive*.

The sample with these primitive objects, which can be only moved, demonstrates a lot of things that are important and used for all the types of moving / resizing even for much more complicated objects. Let's look at some important details.

Default and non-default parameters

The cover for any object is organized as a set of nodes. Nodes can be of three different shapes, so there are three big groups of the `CoverNode` constructors. The parameters, which must be always declared on the node's construction, include identification number, position and sizes. Other parameters often get the default values, but can be declared either during the construction or later. Of the three types of nodes, the polygons are often used to move the objects around the screen, so their default cursor is `Cursors.SizeAll;` nodes of two other forms are usually used for resizing or reconfiguring and their default cursor is `Cursors.Hand`. Because here all three types of nodes are used for moving the objects, I decided to keep their cursors consistent and changed the default parameters in two classes.

```
public override void DefineCover ()
{
    CoverNode node = new CoverNode (0, center, radius, Cursors .SizeAll);
    cover = new Cover (new CoverNode [] { node });
```

Deciding on the buttons to move an object

The `MoveNode()` method allows to specify, by which button each node can be moved. For two of the three primitive classes the possibility of moving only with the left button is declared.

```
public override bool MoveNode (int i, int dx, int dy, …, MouseButtons catcher)
{
    bool bRet = false;
    if (catcher == MouseButtons .Left) {
        Move (dx, dy);
```

For the `PrimitiveStrip` class this limitation is not mentioned; objects of this class can be moved by any button.

```
public override bool MoveNode (int i, int dx, int dy, …, MouseButtons catcher)
{
    Move (dx, dy);
```

The order of objects

In order to become moveable, all objects must be registered in the mover's queue. When the objects are moved around the screen, they can overlap. When two or more objects share the same part of the screen and you want to pickup some element at the point of overlapping, your expectation would be that the object on top has to be caught. Mover doesn't know anything about the drawing of objects and their appearance at the screen; mover makes the decision about the object to catch only according with their order in its queue, so it's the developer's responsibility to draw the objects and to place them into the mover's queue in appropriate order. The elements, viewed on top of others, must precede them in the mover's queue.

There is the strict rule enforced by the system: all the controls are always shown atop all the graphical objects. Thus we receive the rule 1.

**Rule 1.** All the controls must precede all the graphical objects in the mover's queue.

There are four graphical objects and one control in the form; this control must be placed at the head of the queue. Here is the sequence of calls to register the objects in the **Form_Nodes.cs**.

```
mover .Add (text);
mover .Add (circle);
mover .Add (rect);
mover .Add (strip);
mover .Insert (0, btnCovers);
```

The second rule links the order of graphical objects in the mover's queue with the order of their drawing and is based on the fact that graphical objects are painted from the bottom layer to the top.

**Rule 2.** The order of drawing objects must be opposite to their order in the mover's queue.

In other words, the drawing of the graphical objects must go from the tail of the mover's queue to the head. In such a way the graphical object, shown on top of others, will always precede them in the mover's queue, and the top one will be always caught by the mover. Exactly what is expected.

```
private void OnPaint (object sender, PaintEventArgs e)
{
    … …
    strip .Draw (grfx);
    rect .Draw (grfx);
    circle .Draw (grfx);
    text .Draw (grfx);
```

# Into the realm of covers

The previous sample with the primitive figures demonstrated only the moving of those figures and nothing else. Two obvious questions arise after that demonstration.

1. How to resize the objects?
2. How to rotate the objects?

The resizing of an object is usually done by moving its border. Either an object can be resized in any direction or the resizing is limited, depends on the purpose of using this object in an application. In any way, the needed part(s) of the border must be covered with a node or several nodes that are responsible for resizing. As a rule, these nodes are relatively narrow; so the resizing can be started "by the border". There is no requirement for any node to be positioned only and strictly inside the object's area. On the contrary, users expect that they can start the resizing in the vicinity of visual border (not more than several pixels aside) on any side of the border line, so the border nodes often represent the strip (invisible!), which has a real (visual) border line of an object as its median.

A lot of objects have a border, consisting of a set of straight lines; the standard technique is to cover each segment of such border either with a narrow rectangular node or a strip node. Let's begin the exploration of resizing with a simple rectangle.

## Resizing of rectangles

On opening the **Form_Rectangles.cs** (menu position *Nodes and covers – Rectangles*), you see four colored rectangles, which demonstrate four different types of resizing. Visually the differences become obvious only on switching ON the cover visualization (**figure 3**); without those covers in view, the rectangles do not give you any visual tip on how they can be changed. Though there is another prompt – the changing of the mouse cursor, when the mouse crosses any area of possible resizing or moving an object.

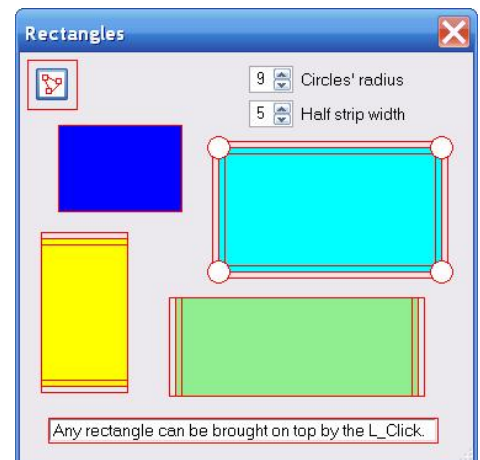

**Fig.3** Resizing of rectangles

The rectangular shape is so popular among the screen objects that a cover for such an object can be constructed by using one of the standard `Cover`'s constructors.

```
public override void DefineCover ()
{
    cover = new Cover (rc, resize, radius, halfstrip);
}
```

The `resize` parameter in this constructor belongs to the `Resizing` enumeration

```
enum Resizing { None, NS, WE, Any };
```

This parameter totally defines the number of nodes in the cover, their types, and order.

- For a nonresizable rectangle (the Blue one on the picture), there is a single node, covering the whole rectangle.

- If a rectangle is resizable only in one direction (Yellow and Green rectangles from the picture), then the two appropriate opposite sides are covered with the two narrow rectangles and then comes the same big node.

- For a fully resizable rectangle (the Cyan one), there are for small circular nodes in the corners, then the rectangular nodes on borders, and then the big one, covering the whole area.

The nodes overlap with each other. At the point of their overlapping, the resizing to be started is decided by the first of those nodes in the cover's array. It's more difficult to find the small node with the mouse than the big one (remember that usually the covers are not shown), so the smaller nodes are usually placed first in this array. That's why for the fully resizable rectangle I decided about such order of nodes: circular nodes from the corners, then the covering of the borders, and then the big node for the whole area.

There are two controls in this form, which allow to change the sizes of the nodes and see, how the easiness of the resizing depends on those values. These controls with the comments can be easily moved around the screen by grabbing them anywhere near those controls or at any point of their comments. This is an object of the `LinkedRectangles` class – one of the classes, widely used for design of the user-driven applications. There will be much more on this and other similar classes further on.

Seven moveable objects in the **Form_Rectangles.cs** are initially included into the mover's queue in such an order:

1. The button to switch the visualization of covers ON / OFF.

2. The `LinkedRectangles` object to change the parameters of the nodes. Both controls with their comments can move only synchronously and never change their relative positions, so it is a single object for the mover.

3. Information (a `TextM` object).

4. Four rectangles (objects of the `RectangleStandard` class) in such an order: blue, yellow, green, and cyan.

The painting of the graphical objects is organized in the opposite order to their positions in the mover's queue; this is according with the rules, declared in the previous chapter. But there is no direct mentioning of those four rectangles inside the `OnPaint()` method of this form. Instead I decided to demonstrate, how the object, to be painted, is obtained with one of the mover's properties.

```
void OnPaint (object sender, PaintEventArgs e)
{
    GraphicalObject grobj;
    for (int i = mover .Count - 1; i >= 0; i--) {
        grobj = mover [i] .Source;
        if (grobj is RectangleStandard) {
            (grobj as RectangleStandard) .Draw (grfx);
        }
    }
```

This was done, because the order of rectangles can be changed by the left click on any of them.

```
void OnMouseUp (object sender, MouseEventArgs e)
{
    if (mover .Release ())
    {
        if (e .Button == MouseButtons .Left   &&
            Auxi_Geometry .Distance (ptMouse_Down, e .Location) <= 3   &&
            mover [mover .WasCaughtObject] .Source is RectangleStandard)
        {
            PopupRectangle ();
        }
    }
```

In the applications, consisting of the moveable / resizable objects, those objects often overlap, so at any moment the needed object from underneath is brought to the top by a simple click. When you click an object with the left button, there is always a question of your intention: was it an ordinary move of the object, or did you decide to bring it on top? In all my applications, the decision between these two cases is based on the distance between the two points, where the mouse was pressed and released. If the distance is really small, then it is considered not a move of an object, but the intention to bring this object on top. On clicking the right button, the similar decision is made between the rotation of an object and calling the context menu.

To bring the clicked rectangle on top, the order of rectangles in the mover's queue is changed, and they are repainted according with the new order.

```
private void PopupRectangle ()
{
    int jCur = mover .WasCaughtObject;
    while (jCur > 0  &&  mover [jCur - 1] .Source is RectangleStandard)
    {
        mover .Reverse (jCur - 1, 2);
        jCur--;
    }
    Invalidate ();
```

## Resizing of polygons and a bit of reconfiguring

The resizing of rectangles doesn't look like a very complicated problem. Let's make one more step, now into the county of polygons, and see, what is different with those figures that can be much more interesting in view. **Form_Polygons.cs** (menu position *Nodes and covers – Polygons*) allows to analyse the behaviour of four different classes. All these classes include in their names the abbreviation **RsRt**, which means Resize-and-Rotate. As their names indicate, all these objects can be involved in rotation, but the time for it will come later; no rotation is demonstrated in this form

- The first class – `RectangleRsRt` – still represents the rectangles. This class is similar to what was already shown in the previous sample with a fully resizable rectangle.

- The second class – `RegularPolygonRsRT` – represents the regular polygons.

- The third class – `PerforatedPolygonRsRt` – is also a regular polygon, but each object has a hole inside; the hole's border has the same form as the outer edge of the polygon.

- The `ChatoyantPolygonRsRt` class represents polygons that can be not only resized, but reconfigured also. Such object has a central point and a set of apexes; apexes are linked into an infinitive loop. An object is composed as a set of triangles; each triangle is based on the central point and two consecutive apexes. All apexes and the central point can be moved individually; in such way a reconfiguring of the polygon is organized. Moving of any connection between two consecutive apexes starts the resizing of the whole object. If you try to move any apex or side of such polygon, you'll understand all these things in shorter time than takes the reading of these several lines. I often initialize an object of the `ChatoyantPolygonRsRt` class in the form of a regular polygon, because the calculation of the apexes in such case is easy; then such polygon can be transformed with a mouse into an unpredictable shape. These objects are also associated with an array of colors, while the objects of the three previous classes are unicoloured.

**Figure 4** can give you some idea of what can be done with the polygons in this form. As always, there are no nonmoveable elements in this form, so the two individual buttons, information text or those three groups that can be seen at the left, all of them can be moved around the screen.

A single button is enough to add the new `RectangleRsRt` object to the collection in the form, but other three classes need some additional parameters, so the objects of these classes are added with the help of those three groups. The groups belong to the `Group` class – another class for synchronous moving of several objects; this class will be discussed a bit later.

We made only the second step into the realm of covers, but we already received a nontrivial collection of moveable objects. Their main features are.

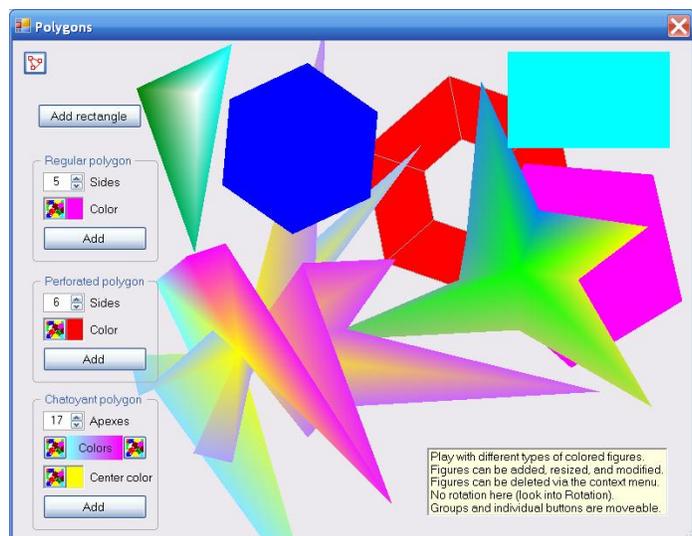

**Fig.4  Form_Polygons.cs**

- Objects of different classes are involved in movements. It is obvious that this number of classes is not limited; any other classes can be added.

- There are no restrictions on the number of moveable objects in the form. This number is not predetermined by the developer of the program, but can be changed at any moment by the user. Objects are added and deleted at any moment.

- Sizes and locations of the objects are not predetermined also. These things are also decided by the users only.

- The whole system of objects can be considered as a multilevel space with each object occupying its own level. The size and position of any object can be changed without any influence on other inhabitants of this world. (There is a technique to prevent the overlapping of the objects, if such a limitation is needed. I don't want to explain this mechanism in this paper, but it is explained in the **Moveable_Resizable_Objects.doc**, which is available at www.sourceforge.net). Any object can be easily (one click) brought atop all others; with the context menu, the level of any object can be changed up or down, thus changing the appearance of objects on the screen.

Though we are dealing here only with the samples of the purely geometrical objects, the mentioned set of features is essential to all the user-driven applications. Further on you'll see that exactly the same features are observed in all the samples from now on; the second article will demonstrate the same features in the most complicated scientific and financial applications. If you combine these features together, then you'll get the main design principle of all the user-driven programs.

> The user-driven application do not determines the rules and limitation of dealing with the designed screen objects. This application is an instrument, with which user can bring up to the screen any number of needed objects and deal with them in any way he decides. The application must give an easy way to add, delete or modify all those objects in order to provide the most efficient way of dealing with them for everyone.

Now let's look at the way of implementing these features in the case of polygons.

Standard rules of cover design.

The cover of any object must be designed in such a way, so that an object can be moved by any inner point and resized by any border point. The inner area of the objects from this form is always covered by a set of polygon nodes. The number of apexes in those polygons can vary, but all the polygons must be convex at any moment. In the `PerforatedPolygonRsRt` class, the division of any object into several polygons is shown by drawing the edges between them (see **figure 4**). The `ChatoyantPolygonRsRt` objects look like regular polygons only at the moment of design; later on they can be quickly turned into the nonconvex polygons, so from the beginning their cover includes a set of triangles, which are later transformed, but always into other triangles, so each of these pieces continues to be a convex polygon.

The border of any object is covered by a set of strips, connecting the neighbouring apexes. This technique is used for any type of objects, except those that have the curved borders. For them the special N-node covers are used; these covers are discussed a bit later.

The strips on the borders are used for resizing (scaling) of the objects. If an object needs reconfiguring, then it is always done by changing the relative positions of its parts. To achieve this, the parts that can change their positions in relation to others, are covered by the individually moveable nodes. If the reconfiguring has to be started by some obvious point of an object (for example, an apex), then this point is covered by a small node. The small circular nodes are often used for such purposes, but the nodes for reconfiguring can be of any shape and any size. Of the mentioned four classes, only the objects of the `ChatoyantPolygonRsRt` class can be reconfigured; all their apexes and the central point are covered with such circular nodes.

Here is the implementation of these rules in the design of cover for the `RegularPolygonRsRT` class.

```
public override void DefineCover ()
{
    CoverNode [] nodes;
    if (bResize) {
        // first the strips along the sides; the last one is the convex polygon
        nodes = new CoverNode [nApexes + 1];
        PointF [] pts = Apexes;
        for (int i = 0; i < nApexes; i++) {
            nodes [i] = new CoverNode (i, pts [i], pts [(i + 1) % nApexes], 5);
        }
        nodes [nApexes] = new CoverNode (nApexes, Apexes);
    }
    else {
        nodes = new CoverNode [] { new CoverNode (0, Apexes) };
    }
    cover = new Cover (nodes);
```

Pay attention to two things.

- The same object can be turned from resizable into nonresizable and vice verse at any moment, while an application is running.  For example, the user needs to change the sizes of the object and then fix them in order to prevent the further accidental change.  This can be done, for example, via the context menu.  (I am using this technique in different programs.)  On such a change, only the cover of the object is changed, but even its appearance on the screen is not, so the application continues its run without problems.  The cover for identical, but not resizable object, is much simpler, as the whole set of nodes to cover the border is not needed.

- There is no indication of possible rotation in this piece of code.  The cover does not depends on either the object is going to be rotated or not.  All the objects in this form are not involved in rotation, but will be rotated in further samples; this will not demand any changes in their covers.  The possibility of rotation is mentioned in the `MoveNode()` method of the class, because the change in rotation status can be decided by the user exactly in the same easy way, as the possibility of resizing.

Identification of objects.

The moving, resizing, and modifying of objects or changing their order depend on the objects' identification.  At any moment, when an object is selected for any of these operations, there must be an absolutely reliable identification of what really is supposed to move, change or disappear.  The first level of identification is provided with the unique ID that each moveable object receives at the moment of construction.  So the first code line in the constructor for any of these objects looks identical and provides the new object with the unique ID.

```
public RegularPolygonRsRt (PointF ptC, …)
{
    id = Auxi_Common .UniqueID;
```

In the complex objects, their parts often can be involved both in synchronous movements with other parts (the movement of the whole object) and in individual movements.  This means that not only the "parent" object is registered in the mover's queue, but all these parts receive their personal places in the same queue.  To decipher the whole chain of relations between the linked objects even if some smaller part was clicked, those parts have to contain the ID of their parent; one level up is enough, as in the same way you can track the relations from level to level.

For example, the object of the `BarChart` class, which can be seen at **figure 1**, consists of the main plotting area with a couple of scales; each scale can be linked with an arbitrary number of comments.  Each comment, which can be moved and rotated individually, receives its personal identification number.  When the new comment is added either to a scale or a main area, it receives also the "parent" ID (either from scale or from the plotting area).  Each scale has its personal ID, but also keeps the ID of the plotting area, with which it is associated.

But this is one form of identification.  Another type of information is provided by the mover; there is an overwhelming amount of information.  Whenever an object is caught by the mover, you can get the information about the type of this object and, for example, start the needed resizing.

```
private void OnMouseDown (object sender, MouseEventArgs e)
{
    ptMouse_Down = e .Location;
    if (mover .Catch (e .Location, e .Button))
    {
        if (e .Button == MouseButtons .Left)
        {
            StartResizing (e .Location);
```

The mover's information is used for the moving of objects

```
private void OnMouseMove (object sender, MouseEventArgs e)
{
    if (mover .Move (e .Location))
    {
        Invalidate ();
        ControlsCausedUpdate ();
```

The mover's information is very valuable even after an object was released, because by analyzing it the next action is decided.

```csharp
        private void OnMouseUp (object sender, MouseEventArgs e)
        {
            int iWasCaught;
            if (mover .Release (out iWasCaught))
            {
                GraphicalObject grobj = mover [iWasCaught] .Source;
                double dist = Auxi_Geometry .Distance (ptMouse_Down, e .Location);
                if (dist <= 3 && grobj is ElementRsRt)
                {
                    if (e .Button == MouseButtons .Left)
                    {
                        PopupFigure (grobj .ID);
                    }
                }
```

Mover can give out not only the position of the object in its queue, but also the number of node, by which it was caught, and the shape of this node. In some cases it's easier to write the code, based on this information, than only on the order of an element. For example, all three types of polygons in the **Form_Polygons.cs** can be resized, but to calculate the correct coefficient of scaling at the moment, when the resizing is started, each class of polygons receives different information. You can see the difference inside the `StartResizing()` method, from which the scaling of the caught object is called with different parameters.

- For the `RegularPolygonRsRT` it's enough to have only the point, where the resizing started.

- For the `PerforatedPolygonRsRT` in addition to the point, the number of caught node is needed; based on this number the decision is made to change the inner or outer border of the object.

- For the `ChatoyantPolygonRsRT` the form of the caught node is provided; if it is a strip node, then the resizing of the polygon is started.

But mover can provide not only the information about an object that is caught, moved or just released. The mover sniffs everything, which is below the mouse, when the mouse is idly moved across the screen without any button pressed. You can get the object below with the `mover.SensedObject`, and then get all the information about this object. (In this situation a mover is like a shark, which is not hungry at the moment and lazy to attack, but watches carefully and keeps track of everything that is going around.) Mover can also provide the same type of information (object, node, node's shape, and so on) not only for the mouse position, but for any point, occupied by any moveable object.

The order of objects.

The order of objects on the screen is regulated by the law, which was declared once and forever by Microsoft and several other barons (any questions about the tyranny?). We are in the kingdom (liberals are banned from here), so controls first (aristocrats!), then all those graphical objects (peasants…). The rules of placing the moveable objects in correct order into the mover's queue and drawing them on the screen, those rules that were explained earlier, are only the consequences of using this law. In the **Form_Polygons.cs** the implementation of those rules means such an order of objects in the mover's queue:

1. The small button to switch the covers' visualization ON / OFF.
2. The button to add new rectangles.
3. Three groups of controls for adding chatoyant, perforated, and regular polygons (in the mentioned order).
4. Information text.
5. All the colored figures.

The first four groups are never changed, but the fifth group can be reorganized at any moment by adding a new figure (it goes to the head of the group), deleting any figure, or moving any existing figure to any position inside this group. The positions of elements in the mover's queue can be changed in an ordinary way by calling the appropriate methods, but I always prefer to call the same `RenewMover()` method, which will set the correct mover's queue at any moment.

```csharp
        void RenewMover ()
        {
            mover .Clear ();
            mover .Insert (0, groupRegular);
            mover .Insert (0, groupPerforated);
```

```
        mover .Insert (0, groupChatoyant);
        mover .Insert (0, btnAddRectangle);
        mover .Insert (0, btnCovers);
        mover .Add (info);
        for (int i = 0; i < elements .Count; i++) {
            mover .Add (elements [i]);
        }
    }
}
```

According with the **rule 2**, the objects' drawing is organized in the opposite order to their placement in the mover's queue.

## N-node covers

Not all the objects that need resizing have the border consisting of the straight lines. Some of them have curved borders, but they still need to be resized by any border point. What is the solution? I call it the *N-node covers*.

In this type of covers, when some part of the border can't be covered with a thin prolonged node (polygon or strip), then it is covered with a set of small nodes. Usually the number of such nodes can be big enough, but it's not a problem, as they are identical in behaviour, so the MoveNode() method of such an object doesn't require a huge amount of code. For some time, I was using in such cases only the small circular nodes. To organize on the border line a sensitive strip, consisting of small circular nodes, the neighbouring circles have to be positioned not side by side, but with a small overlapping. A set of small overlapping circular nodes is only one of the possible solutions. There are no regulations on the number of such nodes, their shapes or the possibility of overlapping. Everything depends on the developer's ideas. The only thing that is required, that this set of small nodes covers the border without any gaps and provides the needed resizing of an object.

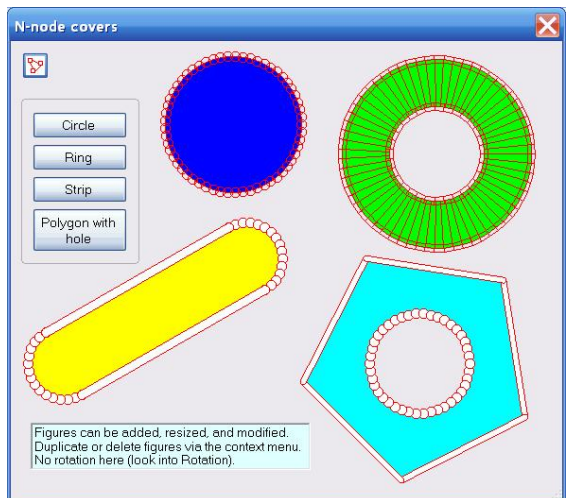

**Fig.5**  Different classes with the N-node covers

Let's look at some samples, which use the N-node covers. I am deliberately going to demonstrate the use of different small nodes for these samples. The two things, which can change, when you switch from one type of small nodes to another, are their number and the use of overlapping; everything else is exactly the same.

**Figure 5** shows the **Form_NnodeCovers.cs**, in which objects of different classes can be moved and resized. The covers of these objects are shown at the picture, so the explanation will be easier.

The cover for the circles (`CircleRsRt` class) consists of one big circle, covering nearly the whole area, and a set of small circular nodes, overlapping each other along the border.

The cover for the rings (`RingRsRt` class) consists of three sets of nodes: first the small polygon nodes, covering the outer border, then another set of small polygons, covering the inner border, and then a set of long narrow polygons, covering the area of the ring. All these polygons have the shape of a trapeze, so there are no gaps between them.

The cover for the strips (`StripRsRt` class) consists of three different parts.

1. Two strips cover the long border lines; these nodes are used to change the width of the strip.

2. Two sets of small circular overlapping nodes cover the curved parts of the border; these nodes are used to change the length of the strip.

3. Big strip node, covering the whole body; the node is used to move the strip.

The cover for the polygons with the circular hole inside (`CircleInsidePolyRsRt` class) consists of four parts: a set of small circular nodes on the inner border, a set of strips on the outer border, one circular node to cover the hole, and one big regular polygon, covering the whole area. This cover has an interesting feature, but we'll return to it a bit later.

Is there anything common in all these covers and at the same time really new, which differs the N-node covers from others? Yes, the changing number of nodes in the cover, when an object is resized. This happens nearly all the time, when the length of the curved border is changed. When the border of an object consists of the straight lines and is covered by the strips or rectangles, then the resizing of such object doesn't change the number of nodes, but only the size (length) of some of them. When the border is covered by a set of nonchangeable nodes, then the change in the border's length requires another number of such nodes to cover it without gaps. There is a very interesting and important consequence of this changing number of nodes.

**The rule of N-node covers.** The `DefineCover()` method cannot be called from inside the `MoveNode()` method, and though the cover has to be changed, because the sizes of an object have changed, but the call to the `DefineCover()` method must be postponed until the release of an object. And before this call, the new number of nodes must be calculated.

For all the previously demonstrated objects with the possibility of resizing, there were no restrictions on calling the `DefineCover()` method on any change of the sizes. It is done, for example, in all the classes for different polygons. When the number of nodes in the cover is fixed for the whole class and does not depend on the size of the particular object, then the `DefineCover()` method can be used at any moment.

When an object is caught by the mover, it is caught by one or another node of the cover. The translation of the node's movement into one or another type of the object's movement is done by the `MoveNode()` method, in which the exact type of movement (forward movement of the whole object, resizing or reconfiguring) is usually determined according with the identification number of the caught node. If the number of nodes is changed during the period of time, when an object is grabbed by the mover, then this identification number can move into the range of nodes that react differently. The result can be absolutely different from what you expect.

For example, I have already mentioned the order of nodes in the `RingRsRt` class: first the nodes on the outer border, then the nodes on the inner border, and then the nodes for the whole area. Suppose that you pressed the inner border and began to squeeze the hole. With the diminishing hole, less and less nodes are needed to cover the inner border. If you would be constantly changing the cover according with the changing size, then at some moment the same number of the caught node will move into the set of nodes that cover the whole ring's area, and instead of squeezing the hole, you'll see the movement of the ring. It is definitely not the expected thing. The rule of N-node covers prevents you from running into such a situation.

The redesign of cover for a previously caught object is done, when this object is released. At this moment the new number of needed nodes is calculated according with the new sizes and the `DefineCover()` method is called.

```
private void OnMouseUp (object sender, MouseEventArgs e)
{
    int iWasCaught;
    if (mover .Release (out iWasCaught)) {
        GraphicalObject grobj = mover [iWasCaught] .Source;
        if (e .Button == MouseButtons .Left)
        {
            if (grobj is ElementRsRt)
            {
                RedefineCover (iWasCaught);
            }
```

## Transparent nodes

I have mentioned before that the cover for the polygons with the circular hole inside (`CircleInsidePolyRsRt` class) has some very interesting feature, now it's the time to look at it more carefully.

The comparison of a ring and a regular polygon with a hole shows some similarity in the general shape as both have the circular hole, but their covers demonstrate a strange difference in their design. The ring's area is covered by a significant number of polygons, while the area of the regular polygon doesn't show such nodes; yet both objects can be moved by any inner point. Certainly, I could easily design both covers in the similar way, but I deliberately made them different to show two ways of design.

The covering of the whole area with a set of nodes is a standard way to make an object moveable by any inner point; this technique was used, for example, in the `ChatoyantPolygonRsRt` class. When you have some nontrivial area (usually it means nonrectangular area, or some holes inside), then you'll need to write more code for the calculations of all those nodes. While introducing the idea of nodes at the beginning of this article, I mentioned that each of them has a parameter, characterizing its ability for individual movement. The value of this parameter is a member of the `MovementFreedom` enumeration.

The `Transparent` member of this enumeration can be very helpful in some situations. When the mover sees the node with such parameter, it not only ignores this node, but also skips all further nodes of the same object. Beginning from this node, an object becomes transparent for the mover and the mover looks for something farther on in its queue to grab and move. Two facts contribute to an extraordinary importance and, at the first glance, a strange use of transparent nodes: the invisibility of nodes and nonequivalence of the object's area and its cover's area.

For the majority of objects, these two areas can be equivalent or nearly the same. To make an object moveable by any inner point, its whole area must be covered by some set of nodes. If an object is moveable, but not resizable, then the sensitive area is limited by the object's area, so the cover's area can be equivalent to the object's area. When an object is resizable, then the resizing is usually done by the border points, in which case the border is covered by the sensitive strip, so the cover's area becomes slightly wider, than the object itself. All this is used in the simple and most obvious cases; for the complicated cases of nontrivial areas only the transparent nodes can help.

While making the decision about the possibility of catching any object, the mover checks the objects from its queue; the cover of each object is analysed node after node, according with their order in the cover. When the first node, containing the mouse point, is found, then there are several possible reactions, depending on the `MovementFreedom` parameter of this node:

- If it is `MovementFreedom.None`, then the mover can't do anything with it. At the same time the object blocks all other objects, which might lie underneath. The analysis is over, try another point.

- If it is `MovementFreedom.Freeze`, then the object under the mouse can't be moved by this point, but it is recognized by the mover as any other object, so, for example, the context menu can be easily called for it.

- If it is `MovementFreedom.All`, `.NS`, or `.WE`, then the real possibility of movement is decided by the `MoveNode()` method of this object, according with the node's number (or shape) and the movements restrictions, if there are any.

- If it is `MovementFreedom.Transparent`, then mover skips this and all other nodes of the same object and continues the analysis of the situation from the next object in its queue.

In some situations, the use of transparent nodes allows to change the covers with a lot of nodes (and a lot of their calculations) into really simple. For example, look at the possible transformation of the ring's cover. All the small nodes for resizing, which are placed on both borders, will be not changed, but instead of a set of polygons, covering the ring's area, two circular nodes can be used.

```
nodes [k] = new CoverNode (k, center, rInner, MovementFreedom .Transparent);
nodes [k + 1] = new CoverNode (k + 1, center, rOuter, Cursors .SizeAll);
```

The case of a ring is really simple; but there are objects, for which the use of a standard technique with a lot of nodes, covering the object's area is either extremely difficult, or nearly impossible. When I first designed the next sample, I called it *SwissCheese*, but because no Swiss cheese has holes in the form of polygons, I changed the name to its current variant. **Figure 6** demonstrates the view of the **Form_AreaWithHoles.cs** (menu position *Nodes and covers – Transparent nodes*). Each rectangular area initially has a number of holes of different shape (circles and polygons); those holes can be closed by the moveable / resizable / rotatable figures of the appropriate shape, in which case the hole disappears and the area's cover is changed. Such a nontrivial area has a very simple cover, consisting of a single transparent node for each hole and a single node for the whole area. Regardless of the number and form of the holes, the number of nodes is always (nHoles + 1).

```
public override void DefineCover ()
{
    CoverNode [] nodes = new CoverNode [holes .Count + 1];
    for (int i = 0; i < holes .Count; i++)
    {
        if (holes [i] .ApexesNumber == 0)
        {
            nodes [i] = new CoverNode (i, holes [i] .Center,
                                   Convert .ToInt32 (holes [i] .Radius),
                            MovementFreedom .Transparent, Cursors .Default);
        } else
        {
            nodes [i] = new CoverNode (i, holes [i] .Apexes,
                              MovementFreedom .Transparent, Cursors .Default);
        }
    }
    nodes [holes .Count] = new CoverNode (holes .Count, rc);
    cover = new Cover (nodes);
}
```

Pay attention that in case of transparent nodes their order in the cover is extremely important. No node can affect any other node, which is positioned ahead of him in the cover. For all nontransparent nodes it also doesn't matter, what is placed behind it in the cover, because the reaction (possibility of movement) is determined only by the first node, on which the mouse is pressed.  The transparent nodes are the only exception of this simple mechanism: when such node is pressed, it doesn't allow to make any decision on the possible movement, but orders the mover to skip the remaining part of the cover from this object and look for another one.  Ridiculous behaviour, but very interesting and in some cases very helpful.  Certainly, there is no limitation on the number of objects that can be skipped, if the mouse was pressed on their transparent nodes, so in the shown form it is possible to move the lower area, if it was grabbed through the holes of all the upper areas.

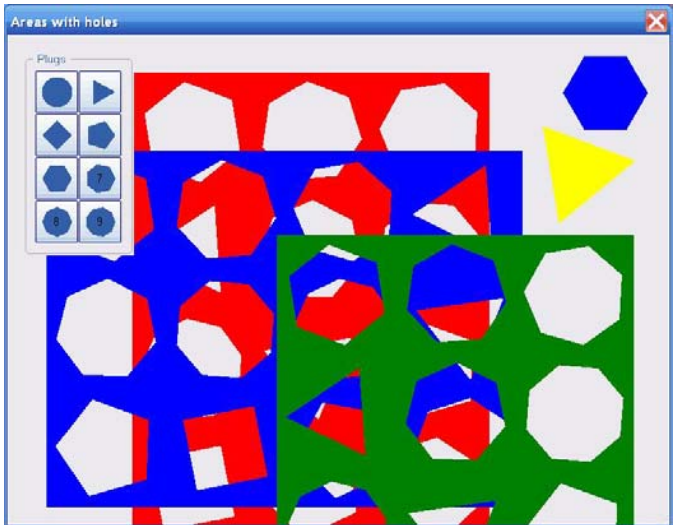

**Fig.6**  Areas with holes; the holes are covered by the transparent nodes

This sample of area with holes still represents the case, when the object's area is nearly the same as the cover's area, but the transparent nodes often organize a situation, when these two areas become absolutely different.  Consider the case of a rectangle with the slightly concave sides: the cover is very simple (a rectangle plus four big circles), but the cover's area is much-much bigger than the object's area, as circles with big radiuses are used to organize the concave sides.

There is one more very interesting object in the **Form_AreaWithHoles.cs**: the group of buttons that are responsible for adding the new figures to the form.  Each button works in a standard way: click the button – receive the new figure.  It's the behaviour of the group, which is not normal.  Certainly, the whole group can be moved around the screen and placed anywhere you want: simply press at the frame or anywhere inside and move the group.  This is not an extraordinary thing, as there is not a single unmoveable object in this demo application.

But each button inside the group can be resized and moved individually; the frame will adjust itself so, as to be around the whole set of the inner elements regardless of their sizes and positions.  This is an object of the `ElasticGroup` class – another class, which is used to design the user-driven applications.  The time to discuss this class and the consequences of using it for the design of applications will come further on in this article and especially in the second article.

We looked at the forward movement, resizing, and reconfiguring of the objects, now it's time to look at their rotation.

## Rotation

All the previously described classes with the abbreviation **Rt** in the names can be involved in rotation, so they don't need any changes at all.  This is a common rule not only for these classes, but for all the moveable objects: the design of their covers does not depend on either those objects are going to be rotated or not.  The covers are designed to provide the needed movements; the start of the forward movement and rotation are distinguished not by the touched place of an object, but by the used mouse button. (Certainly, it can be organized in a different way, when the places to start the forward movement or rotation would be different, but, from my point of view, it would be a wrong design, because then users would have to know and remember the difference between those areas.  It's much better, when any object can be moved and rotated by any inner point.  This will not demand from the users any extra knowledge about the screen objects, with which they deal in this program or others.)

Any object, which can be involved in rotation, has to have a basic angle among its parameters.  Usually one angle is enough, because during the rotation the relative angles between the parts are not changed, so the positions of all the basic points are calculated according with this angle and the sizes.

During the rotation, all the object's points (with the exception of the rotation center, if it is inside the object) change their positions, so the positions of all the points, on which the drawing of an object is based, must be recalculated.  The obvious place to do it is inside the `MoveNode()` method in the part, which is associated with rotation.  For example, here is this part of code for the `StripRsRt` class.

```
override bool MoveNode (int i, int dx, int dy, Point ptM, MouseButtons catcher)
{
    … …
    else if (catcher == MouseButtons .Right  &&  bRotate)
    {
        double angleMouse = Auxi_Geometry .Line_Angle (center, ptM);
        angle = angleMouse - compensation;
```

```
            ptC0 = Auxi_Geometry.PointToPoint (center, angle + Math.PI, length / 2);
            ptC1 = Auxi_Geometry .PointToPoint (ptC0, angle, length);
            DefineCover ();
            bRet = true;
        }
```

Several things, which are important for rotation, can be seen here.

1. The number of the caught node is not mentioned in this part of code, because it doesn't matter at all; the rotation of an object can be started by any node.

2. The parameters of linear movement of the mouse (*dx* and *dy*) are not mentioned either; I found it much more reliable to base the rotation on the exact mouse position *ptM*.

3. You can be surprised to see the call to the `DefineCover()` method here, as it is against the rule, which was declared for the objects with the N-node covers. That rule is important only for the process of resizing and reconfiguring, when the number of nodes can be changed. During the rotation, the shape of an object is fixed, so the number of nodes cannot change, and the rule is not important. Certainly, the call to the `DefineCover()` method can be taken out of here, but then the cover must be redefined at the moment, when an object is released after rotation. The `DefineCover()` is never a time consumption method, so it's much easier to do it on any rotation, than to add the special parts of code.

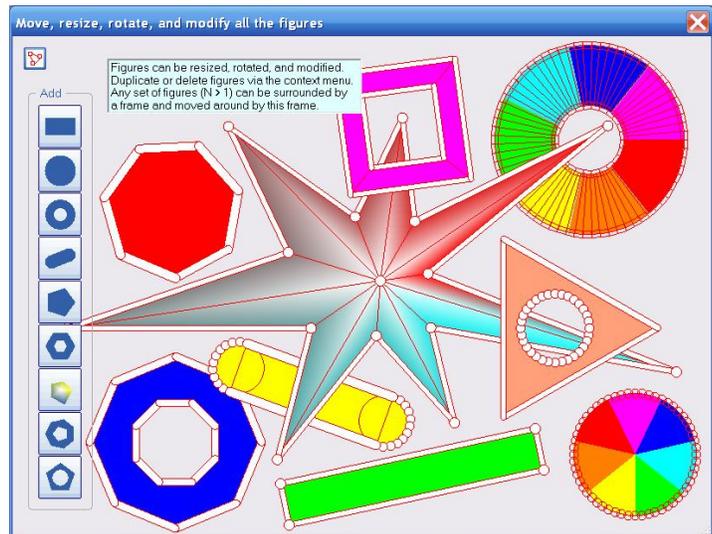

The really interesting thing here is the use of the ***compensation*** parameter, which brings us to the whole process of organizing the rotation.

**Figure 7** shows the view of the **Form_Rotation.cs**, in which any number of objects from different classes can be moved, resized, and rotated. In this figure, the objects are shown together with their covers, so you see the exact situation, which mover has to analyse, while deciding to start any kind of movement. When any object is pressed with the mouse, then the decision on either to start the rotation or any other kind of movement is based on the pressed button.

**Fig.7**  Rotation of different objects

```
        private void OnMouseDown (object sender, MouseEventArgs e)
        {
            ptMouse_Down = e .Location;
            if (mover .Catch (e .Location, e .Button)) {
                if (e .Button == MouseButtons .Left) {
                    StartResizing (e .Location);
                }
                else if (e .Button == MouseButtons .Right) {
                    StartRotation (e .Location);
                }
```

The start of resizing was discussed previously; for some classes it would require additional information either in the form of the node's number or its shape. For rotation, no other information is needed, except the point, where an object is pressed. But each class of objects has its own method, which is called at the start of rotation. The purpose of this method is to calculate this compensation and, maybe, some additional parameters (sizes), which are not going to change during the whole period of rotation. Here is this method for the `StripRsRt` class.

```
        public override void StartRotation (Point ptMouse)
        {
            center = Center ();
            double angleMouse = Auxi_Geometry .Line_Angle (center, ptMouse);
            compensation = Auxi_Common .LimitedRadian (angleMouse - angle);
            length = Auxi_Geometry .Distance (ptC0, ptC1);
        }
```

Each object is characterized by its angle, on which the whole drawing of this object is based. The rotation can be started by pressing an object at any point, so the *compensation* is the difference between the mouse angle and this basic angle at the start of rotation. During the rotation, this difference is not changed, the basic angle is calculated from the changing mouse position by using this fixed compensation, and the whole object turns, as the mouse goes. These calculations are made in that part of the `MoveNode()` method, which is related to rotation; that is where the *compensation* parameter is used (see the code on the previous page).

One interesting aspect of moving the graphical objects is not seen at **figure 7**, though it works in the **Form_Rotation.cs**. The objects in all the previous samples are moved individually. If there are a lot of objects on the screen, then in addition to individual movements, it would be nice to have an easy way of moving the groups of objects. And not only the set of objects that were unified into some kind of group at the design time, but an arbitrary set of objects, which user would like to relocate synchronously. In this form you can select any part of the screen by pressing the left button at any empty place and moving the mouse across the screen. If more then one figure is caught inside this rectangle, then the special frame (of the `SimpleFrame` class) appears on the screen. By moving the frame, all the figures inside are also moved. Selection of an empty area or an area with a single element inside automatically deletes the frame.

The area of this colored frame can be grabbed for moving not by any inner point, but only by the frame itself (the red line); there are also several visible nodes to change its size, which allows to include and exclude the figures from this synchronous movement. The frame will not include into the synchronous movement the controls, which happen to be inside, but this is only because we didn't start looking at controls. In reality, there is a commented piece of code inside the `OnMouseDown()` method. Turn this comment into working code, and the controls will move also, but the discussion for this will come further on.

## Separately and together

All the previously explored objects were moved forward, resized, reconfigured, and rotated individually. Though reconfiguring of the polygons from the `ChatoyantPolygon` class is the result of changing the relative positions of their parts, but these parts can't be looked at as individual objects: an apex or central point of a polygon can't live by itself.

It's a common enough situation, when individual objects are composed into something more complex, in which case those smaller parts can live (move, rotate) by themselves, but at the same time the proposed union is also treated as a single object, in which everything can move and rotate synchronously. Thus we receive a combination of individual and synchronous movements, where the parts have to inform each other about the changes in their positions and sizes. Let's look into the problems of such movements.

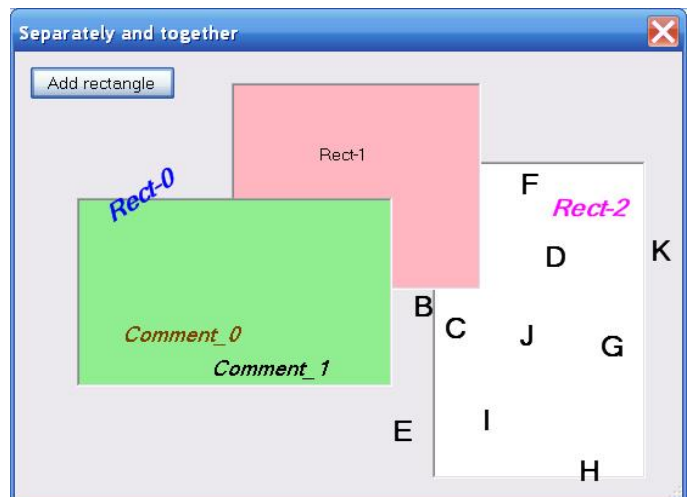

**Fig.8**  Objects, moving individually and synchronously

The **Form_SeparatelyAndTogether.cs** demonstrates the colored rectangles (class `RectangleWithComments`) and the comments (class `CmntToRectangle`), which are associated with them (**figure 8**). As usual, there are no limitations on anything user would like to do with these objects: add, delete, remove, change the order, or change the parameters at any moment. There is a single button to add the new rectangles; everything else is done with a mouse or via a couple of context menus (one for rectangles, another for comments). The individual and synchronous movements of these objects are organized in such a way.

- When any rectangle is moved, then all its comments move synchronously with the rectangle.

- When a rectangle is resized, then each of its comments is relocated, but the new position for each of them depends on either the particular comment was originally inside or outside the "parent" rectangle. (Comment's position is described by its central point.) When a comment is outside the rectangle, then its distance from the rectangle is kept constant regardless of the rectangle's change. If a comment is inside, then its relative position inside the rectangle is kept unchanged.

- Any comment can be moved and rotated freely; such movements have no effect on anyone else.

It looks like a very simple set of rules and a very simple sample, but exactly the same rules are applied to the objects, on which the plotting in very complicated scientific / engineering applications is based. At **figure 1**, several classes of financial plotting are demonstrated; all those classes use the same technique to organize synchronous and individual movements of their parts.

The first and the most important thing in organizing such movements is the correct registering of these complex objects in the mover's queue. Here are several statements that must be considered.

- A rectangle might have an arbitrary number of comments.

- These comments can move individually, so each comment must be registered in the mover's queue individually.

- The comments can be placed anywhere, but they are always shown atop their "parent" rectangle. To be shown above the rectangle, the comments must be painted after this rectangle; then according with the previously declared rule, they must be registered before their parent in the mover's queue.

- The comments can be added and deleted at an arbitrary moment; the request for such actions can come from different places in the code.

Combine these statements together, and it will be obvious that it is unreliable to change the queue manually on any changes in the comments' situation. Such complex objects with individually moveable "children" always have to use a single method, which will guarantee the correct registering, regardless of the components' number. Here is such `IntoMover()` method for the `RectangleWithComments` class; a rectangle will be correctly registered together with all its comments regardless of their number.

```
public void IntoMover (Mover mover, int iPos)
{
    mover .Insert (iPos, this);
    for (int i = comments .Count - 1; i >= 0; i--)
    {
        mover .Insert (iPos, comments [i]);
    }
}
```

The change in the number of moveable objects in the form might happen in different cases: add a new rectangle, delete a rectangle with all its comments, and add or delete any comment. In each case, the `RenewMover()` method is called.

```
void RenewMover ()
{
    mover .Clear ();
    for (int i = rects .Count - 1; i >= 0; i -- )
    {
        rects [i] .IntoMover (mover, 0);
    }
    mover .Insert (0, info);
    mover .Insert (0, btnAddRectangle);
```

These two methods work in pair:

- The `IntoMover()` method guarantees that any rectangle is registered correctly regardless of the number of its comments.

- The `RenewMover()` method guarantees that all the form's objects are registered fully and correctly. Pay attention that the comments are not even mentioned in this method, because their correct registering is hidden inside the `IntoMover()` method.

The `RenewMover()` method is developed for each form with the changing number of moveable / resizable objects. The `IntoMover()` method is designed for each complex class with the individually and synchronously moveable parts.

Mover doesn't know anything about the real objects, but deals only with their covers (in the form of the nodes' array). If any node is caught for moving, this is translated into the `MoveNode()` method of the corresponding object. Mover doesn't know anything about either it is going to be an individual movement or a synchronous one; only the correct method of the caught object is invoked, so the request for synchronous movements of all the comments must be somewhere inside the rectangle's `Move()` and `MoveNode()` methods. It is really there: on any movement of the rectangle, all its comments are informed about the new rectangle area via this method.

```
private void InformRelatedElements ()
{
```

```
            foreach (CmntToRectangle comment in comments)
            {
                comment .ParentRect = rc;
            }
        }
```

Any comment has two instruments to set its position in relation to the parent: the parent's area (rectangle) and the coefficient (in reality, two coefficients) to describe the position in relation to this rectangle. On any movement of any object only one of these parameters is changed; then the second one is used to recalculate the position.

- If the rectangle is moved or resized, then the fixed coefficients are used to calculate the position according with the changed area.
- If a comment is moved, then its new point is used to calculate the new coefficients in relation to the unchanged rectangle.

This is the procedure for rectangles with comments, but exactly the same technique is used for other types of related objects. For example, the circular types of financial graphics, which can be seen at **figure 1**, use two different types of comments: some comments are associated with the whole plot (circle), others with its parts (sectors). The calculation of the specific coefficients is different for each type of comments and depends on the shape of the "parent" object, but the idea is exactly the same.

It also doesn't matter, how many levels of related objects are included into the chain of linked objects. For example, the scientific plots and the bar chart, which can be seen at **figure 1**, include cases of two and three levels of related objects: the main plotting area, the related scales, and the comments, associated either with the scales, or with the main plotting area. The number of levels of objects, included into the synchronous movements, doesn't matter; the individual and synchronous movements in different combinations of all these objects are organized exactly in the same way, as it was shown for those simple rectangles with comments.

Certainly, all these things can't work without the reliable identification of all the involved objects. Each object gets its unique identification number, as was already shown in the case of polygons. That was enough for the case of individual movements; it is also required but not enough for more complex cases.

When any new comment is organized, it receives its personal ID.

```
        public CmntToRectangle (…)
        {
            id = Auxi_Common .UniqueID;
```

When this comment is added to some rectangle, it also gets the parent's ID.

```
        public void AddComment (CmntToRectangle comment)
        {
            comment .ParentID = id;
            comments .Add (comment);
        }
```

That is the thing that provides the identification of all the related objects in the chain, whenever any of them is picked out for moving, deleting or any other change.

One more thing, which you can find, when you start playing with those rectangles. There are two ways to add a new comment to rectangle; both of them can be started via the context menu.

The first one – *Add comment (quick)* – simply adds a new comment with the "New comment" text, using the current `Font` and `ForeColor` of this form. (The font and color of each comment can be changed later via the context menu on this comment.)

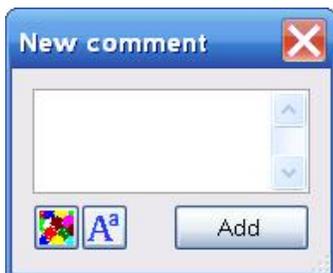

**Fig.9  Form_NewComment.cs**

The second way – *Add comment (custom)* – allows to put any text into the new comment; at the same time the font and the color of this comment can be declared. All these things can be done in the small form, which is opened on clicking this menu line. The **Form_NewComment.cs** is a simple one with only four standard controls inside (**figure 9**), but I have to remind you again that there is not a single unmoveable object in this application. So all the controls in this form are moveable (by their borders), two of them are resizable (corners are the best places to try). Two small buttons are left nonresizable, but it can be changed easily (the explanation starts several lines further on); the whole form can be changed in any possible way you want.

This is the way all the user-driven applications are organized. The purpose of this form is to set the new text, its font and color. This is done regardless of the positions or sizes of all those controls. Then why their positions and sizes must be fixed by the designer? If any user would like to change the view of this form, he can do it according with his own taste. The form works according with its purpose regardless of its view. An unlimited customization.

This form has no graphics and includes only controls; yet it is operated in the similar way to all the previously shown forms, where all graphical objects are moved without any restrictions. But dealing with the controls is slightly different, than working with the graphics. The next part of the article is about making all controls moveable / resizable and working with such moveable controls.

## And controls also

Applications are filled with the elements of two different types: graphical objects and controls. In reality, all the screen elements are graphical. "All animals are equal, but some animals are more equal than others" Nothing can describe the situation with the controls better, than this famous statement.[1] The controls are those "more equal" objects, which makes them absolutely different.

When you design and work with the applications, composed of the fixed controls with the occasionally repainted background, you have no problems with the controls. You simply got used to whatever is provided with them (all their properties and events) and what you can get of them, as a developer. The users also got used to the controls' view throughout the last quarter of a century and demonstrate the classical Pavlov's reflex: if they see the button, they click it. And it really works in such a way, which makes the reflex only stronger.

What none of the users try to do is to move the buttons, for example, to another place, even if they think that another place would be better. Well, I think that some of them tried, when there was no one around to laugh at their attempts, but immediately found out that that was impossible. The reflex was fixed forever: you can click the controls, but not move.

In reality this statement about the controls is absolutely wrong; they are moveable, they are even developed to be moveable and resizable, but these features are closed from the users and occasionally are used by the developers to make an impression. (Like some people in the passed centuries made their living on the ground of knowing, which gradients to through into the flame to produce the colored smoke. Some of those people were burnt for such knowledge, but that is an absolutely different story.) The big part of the popular dynamic layout is based on using the moveability and resizability of the controls.

Controls can be easily moved and resized by using their properties; the demonstration of a button, running away from the approaching mouse, is just a funny sample of using these properties. (Write the code for one mouse event, consisting of several primitive lines, and the users will be amazed with the behaviour of those crazy controls.) But this is an example of how the developers can use these features, because such a running away control is going to move according with the rules, predefined by the developer. I am writing about the mechanism that allows USERS to move controls in any way they would like to do, so it is absolutely different, though I use exactly the same controls' properties. It's the question of who is managing the controls' movements: either the algorithm is absolutely predetermined by the developer (the control is moved for predefined number of pixels in predetermined direction, when the mouse arrives over it), or all the movements are determined only by users, when they would decide to move a control one way or another. I think that users and only users must get the full control of moving or resizing the controls.

### Individually moved controls

The idea of moving and resizing controls by a mouse is exactly the same, as with all the graphical objects, but there is a problem. The standard procedure for the graphical objects is to move them by their inner points and resize by the borders. Unfortunately the whole area of any control is closed for anything new, as controls are designed to use their every inner point for the mouse clicks with already declared purposes. The reflex is so strong that can't be changed. Thus the only chance to move and resize the controls is to use the vicinity of their borders. And as we need both the moving and resizing, then the control's frame is going to be divided between the areas to start moving or resizing.[2]

---

[1] G. Orwell, Animal farm, 1945

[2] There is some inconsistency in using the border of the graphical objects for resizing only, but the border of the controls for both actions. That is one of the consequences of having these "more equal" elements on the screen. The controls can be redesigned and treated like all other objects, but this would be really a big step, though I think it will happen some time in the future. There is absolutely nothing that demands the existence of those controls on the special level. Any one of them can be replaced with the graphical object that looks exactly the same, behave exactly the same, but does not demand the special status. I designed such objects before, other people were doing it, and so this is not something absolutely unique and never heard about. It's impossible for the users to distinguish the current day controls from the similarly designed graphical

Certainly, this frame around a control is used as a cover. It looks a bit strange to have a cover outside an object's area, but there are two remarks to this situation. First, the cover often goes outside the real object; in fact, it goes outside every time, when an object is resizable; the covering of the border makes sensitive an area on both sides of the border. Second, this cover, used for moving / resizing control, belongs not to the control itself, but to a `FramedControl` object, which wraps this control. This object – a frame around the control – is registered in the mover's queue as any other object. The only difference is that the moving / resizing of this frame are directly translated into the moving / resizing of the associated control, making it moveable and resizable. The only question in organizing such a frame is the placement of the nodes for resizing, though several applications use the same technique, so the practice is already known. For example, when you set the size of the proposed control in Visual Studio, you resize it by the corners or by the small squares in the middle of the sides. So, the best and expected places for the nodes to resize any control are obvious, other details depend on the purpose of the needed resizing.

**Form_ControlsIndividual.cs** (menu position *Controls – Moving individually*) demonstrates several individually moveable controls with their covers (**figure 10**). Each control is wrapped in a `FramedControl` object; then this object is registered with the mover.

```
mover .Add (new FramedControl (ctrl, bMove, nodesize, frame));
```

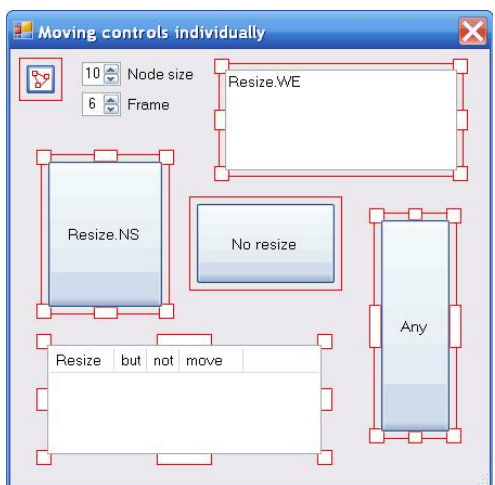

**Fig.10** Individually moved controls and their covers

As you can see, there is no parameter to declare the type of resizing, but the controls in this form can be resized in different ways. The mover determines the type of resizing for a particular control by analyzing its sizes and the values of its `MinimumSize` and `MaximumSize` properties. If those properties are not changed from their default values (0, 0) or set to the size of the control, then the control is nonresizable. If there is a range for one direction only (width or height), then this control is resizable in this direction only; otherwise it is fully resizable.

The `frame` parameter determines the width of the sensitive frame around the borders; by pressing inside this area, the control can be moved. **At figure 10**, the covers are visualized and these sensitive areas around the controls are shown by the big red frames. The width of a frame can't be less than 2 pixels (by default it is 6 pixels), but it can be eliminated, if the second parameter `bMove` is set to `false`; the control for such case is shown in the left bottom corner of this figure. It looks like this control without a frame is resizable, but not moveable. In reality it's not so. There is a funny way to move even this control, but exactly like all the caterpillars move: stretch it in one direction by the small node, then squeeze by the opposite node, stretch again, and so on.

Making everything moveable in the form, consisting exclusively of controls, is really easy. Just register all the controls from the `Controls` collection with the mover and everything will work fine.

```
foreach (Control ctrl in Controls)
{
    mover .Add (new FramedControl (ctrl));  // or  mover .Add (ctrl)
}
```

The code for three mouse events can't be simpler, than it is in this form. In such a way ANY form, containing only the controls, can be turned into moveable / resizable / changeable just in seconds. Are there real cases, when such a simple turn of all the controls in the form into moveable / resizable would work? Certainly, the **Form_NewComment.cs**, which was mentioned a couple of pages back and which is used to add the new comments to the rectangles, is just such a form.

## Groups of controls

Now let's look at the different case in our kingdom of moveable objects. Controls are often used not as stand alone objects, but in groups. For example, when you need to define several parameters of the same object, you often unite a set of needed controls into a group. It's a standard situation that these controls are positioned next to each other, and there is some kind of visual prompt that they are involved in related processes. This group of controls can be positioned either on a `Panel` or in

---

objects; they will not see the difference at all. For programmers, such graphical objects are much better for the forms' design, as they can be of an arbitrary shape. Certainly, this need some work to be done, so that the new graphical "controls" will be as easy in use for design, as our modern day controls, but I think that the word *progress* is a correct one to describe such a work to be done.

a `GroupBox`; the last one can use its title to provide information about the group. Anchoring can be applied to such `Panel` or a GroupBox, anchoring can be applied to the controls inside, so the sizes of those controls can be adjusted to the sizes of the whole form or its font.

Exactly the same thing can be done with the ordinary panels and group boxes in our universe of moveable objects. `Panels` and `GroupBoxes` are the ordinary controls, so they can be registered with the mover and thus moved and resized exactly in the same way, as was shown for the individual controls. It doesn't matter that they contain other controls; the same rules of the dynamic layout will work. It is possible, it is simple, and from my point it is not a good solution. Though the inner area of a `Panel` or a `GroupBox` looks like a part of the form's area (usually they have the same background color), but it is the inner area of a control, so it is forbidden to be used by the mover; both elements can be moved only by their borders.

There can be much better solutions to move and resize the groups of controls; these unions of elements can be based on different ideas. In this article, I'll introduce only three different classes; there are more in the **Test_MoveGraphLibrary** application and the related text.

Three groups in the **Form_ControlsGroups.cs** (menu position *Controls – Groups,* **figure 11**) behave differently. Certainly, you would never put groups with different behaviour into a single form of a real application, but it's not only OK in the demo program, but maybe even better for their comparison. The name of the class, used for each group here, is shown as its title. There is no need to think about the covers for these groups or methods of their moving; the only needed thing is to organize them according with the described rules (they are different for each class). All three groups are moveable by any inner point, but they differ in the way of resizing.

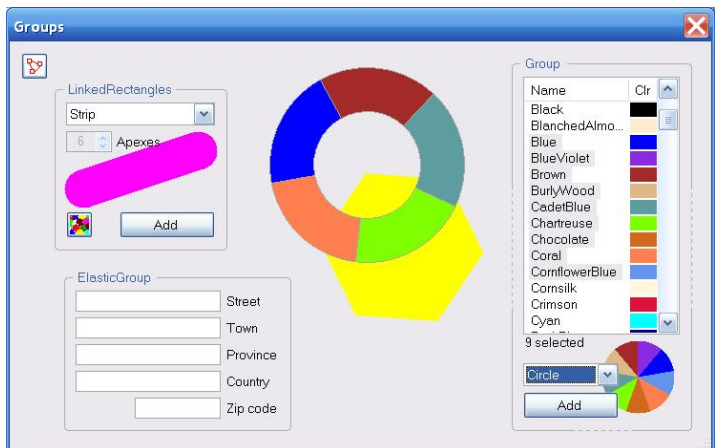

**Fig.11** Different variants of organizing groups of elements

Objects of the **LinkedRectangles** class are not resizable at all. It's a rare situation in the world of moveable / resizable elements, but occasionally it can happen, when required. There are four controls in this group: three of them are used for selecting the type of the new graphical element and its parameters; an additional button is used to add this new element to the screen. The sample of the proposed figure is also shown inside this group.

There can be any number of elements, of which the `LinkedRectangles` object is composed. There can be controls; there can be rectangles with some painting in them. The drawing of a frame is just one of the possibilities; it is not mandatory, so such an object can be easily shown without any frame. There are absolutely no rules for positioning of all those rectangles: they can be placed next to each other, they can be positioned apart, and they can overlap. The order of assembling the rectangles into this object is also not important, as all the nodes of the cover work for one purpose only: to move the whole group. In this form, the `LinkedRectangles` object is organized on the basis of an array of controls; then one more rectangle is added to avoid any nonsensitive gaps inside.

```
Control [] cntrls = new Control [] { comboUnicoloured, numericUD_Apexes,
                                      btnUniColor, btnAddUnicoloured };
lrUnicoloured = new LinkedRectangles (cntrls);
Rectangle rcFrame = Auxi_Geometry .FrameAroundControls (cntrls, spaces);
rcFrame .Inflate (3, 3);
lrUnicoloured .Add (rcFrame, "Area");
```

For the mover, the whole group is a single moveable element, so it can be registered in the easiest way.

```
mover .Add (lrUnicoloured);
```

Objects of the **Group** class are moveable and resizable. The group has a classical view of the group with a frame surrounding the inner area; it is also the classical type of a moveable / resizable object, which can be moved by any inner point and resized by any border (frame) point. Depending on the need, such group can be organized with different types of resizing. There is a visual prompt about the possibility of resizing, as the sides of the frame, which can change their length, have a piece of dashed line in the middle (there is no dashed line at the upper side, if the group has a title).

```
rcFrame = Auxi_Geometry .FrameAroundControls (new Control [] { listColors,
                            comboMulticoloured, btnAddMultiColoured }, spaces);
```

```
groupMulticoloured = new Group (this, rcFrame,
                                new RectRange (…), "Group", MoveGroup);
```

The possibility of resizing the particular group is defined on initialization by the `RectRange` parameter; depending on this parameter, the group can be nonresizable, resizable only in one direction, or in both. The relocation and changing of the sizes of the inner elements are described by another parameter. The moving / resizing of the inner elements happen as the result of the frame's change. These inner elements can't move by themselves independently of the frame's changes, so for the mover this group looks like a single element; thus it is also registered in the easiest way.

```
mover .Add (groupMulticoloured);
```

If the `Group` class can be looked at as the implementation of the dynamic layout for the moveable group, the situation with the **ElasticGroup** class is absolutely different. In this class, the frame's position and size are determined by the set of inner elements, any one of which is moveable and possibly resizable. You move or resize any text box in the shown sample, or you move / rotate any text, associated with these text boxes; the frame watches all the changes and adjusts its position according with all the inner changes.

The `ElasticGroup` object at **figure 11** represents the standard combination of the controls to deal with the address information. Though this order of writing the address is normal for some western countries, it is unnatural for other countries, in which the address is written in the opposite order. With such moveable controls inside the group, their positions can be changed by anyone in seconds. This group of individually moveable inner elements also helps to avoid other annoying things. Some users would like to see the comments on the other side of the text boxes, some would prefer to position those boxes in the different way, give them other sizes, or whatsoever. Users can position / resize the inner elements in any way they want; and in any case they can move the whole group around the screen.

Because the `ElasticGroup` object contains the individually moveable parts, it must be registered by its `IntoMover()` method.

```
groupAddress .IntoMover (mover, 0);
```

The possibilities of organizing moveable / resizable groups of controls are not limited to these three cases. The use of moveable / resizable elements as the basis of construction immediately ends the monopoly of the `GroupBox` class and opens a wide area for new ideas. For example, all these three demonstrated classes use the classical idea of a moveable element, which can be moved by any inner point. This is an excellent rule for graphical objects, but maybe this rule must be changed, when applied to the groups. In couple of minutes each of those samples can be changed in such a way that the groups will be not moved by inner points, but moved and/or resized only by their frames. The division of the frames on areas for moving / resizing will be done exactly in the same way, as was shown for individual controls, so there will be no problems in using such frames for both things. These are some of the suggestions; there can be others. (I have already designed a set of different classes, but decided not to demonstrate them here; as it will only increase the text.)

Several forms in the accompanying application use the `ElasticGroup` objects, but in all these samples the objects of this class are used in the most primitive way. Much more interesting cases of using this class will be described and demonstrated in the second article.

**Technical note.** When you deal only with the graphical objects, then the use of the standard double buffering solves the problem of the screen flickering. When you start to move controls or group of controls, then the system makes the decision about the proper moment for their redrawing, and it is always done with some delay. To solve the problem and improve the whole picture, add a couple of lines into the `OnMouseMove()` method. Throughout the code of these application, it is done with the `ControlsCausedUpdate()` method; depending on the form, the types, which are checked inside this method, can vary, but they include all the classes with the controls that can be moved in the particular form.

```
private void ControlsCausedUpdate ()
{
    if (mover.CaughtSource is ElasticGroup || mover.CaughtSource is FramedControl)
    {
        Update ();
    }
}
```

## Arbitrary grouping

Controls can be moved individually and controls can be organized into groups with different rules for moving. The location and size of these groups can be changed by users, but not the content of each group, which was determined at the design stage. Some applications include a significant number of controls; users can move and resize each of them individually, but in addition they would like to get an easy to use instrument to rearrange the view of such applications more quickly. Not moving one control after another, but uniting any number of them into a group of synchronously moving controls. These groups can not be predetermined by the designer, but has to be organized by any user at any moment. What is the solution? The familiar `ElasticGroup` class.

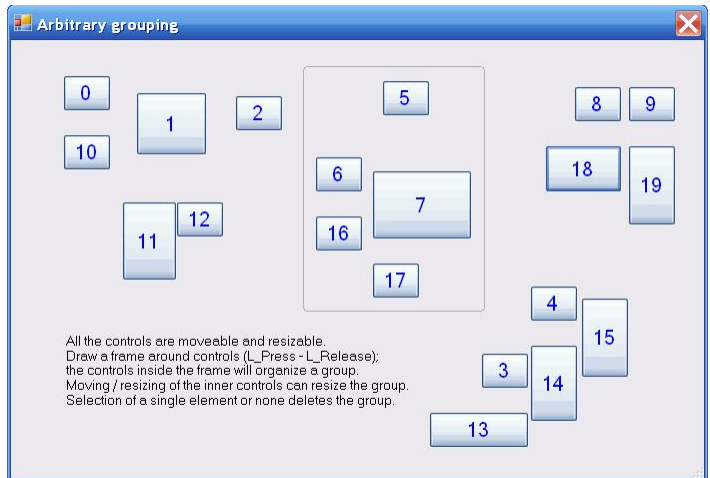

**Fig.12** Form with an arbitrary grouping of the controls

**Figure 12** shows a typical view of the **Form_ArbitraryGrouping.cs** (menu position *Controls – Arbitrary grouping*), in which any control can be moved and resized individually, but at any moment an arbitrary set of controls can be rounded into a group, organizing an object of the `ElasticGroup` class. As was mentioned before, such object with all the inner elements can be moved around the screen by any inner point.

On opening this form, you'll see 20 buttons of the same size, strictly placed in two lines. From that moment you are the only master of these buttons; you can resize and place them in any possible way. If you want to move any group of controls to a new location, you surround them with a temporary frame. This is done in the standard way, which is common for such operations: press the left button at one point, drag the mouse to another point, and release the button. The two points will be the opposite corners of the rectangle. The rounded controls form an `ElasticGroup` object. As any control in the form can be moved and resized individually, then there is no sense in organizing such an object, if there is less than two buttons inside.

Controls inside or outside the frame can be moved and resized individually in exactly the same way. The only difference between them that the inner elements can change the size of the group, as it is an idea of the `ElasticGroup` class. This is a demonstration, which shows, how easy it is to implement such a powerful instrument. I have designed the whole Calculator, based on exactly the same instrument; this Calculator can be seen as a part of the **Test_MoveGraphLibrary** application and as a stand alone application in the mentioned project at www.sourceforge.net. I'll write more about it in the second article, but I think it is obvious that the design of programs on such principle gives to the users absolutely new opportunities in customizing the view of the applications.

# Summary

In my approach to turning the screen objects into moveable / resizable, each object receives an invisible cover. Here are some rules on organizing such covers and on organizing the whole moving / resizing process for the screen objects.

## *Cover summary*

- A cover consists of an arbitrary number of nodes. The minimum number of nodes is 1. A cover may consist of a single node. Number of nodes is unlimited. Covers of a special type (*N-node* covers) may include hundreds of nodes.

- Nodes can be of different shape: circular, strip with semicircles at the ends, and convex polygon. Circular nodes are defined by a center point and radius. Strip nodes are defined by two middle points at the ends of a strip and radius of semicircles; the strip's width is equal to the diameter of semicircles. Convex polygon is defined by the apexes.

- Nodes of a cover may overlap, they can be placed side by side or apart. The order of nodes in the cover can be very important, as nodes are checked for moving according with this order. In the areas of overlapping, moving / resizing of an object is determined by the first selected node.

- Nodes do not duplicate the shape of an object and they are not required to be only inside the object's area. Use of the nodes with the transparent property can make the design of covers for the nontrivial areas much simpler.

- Nodes can be moved individually thus allowing to reconfigure an object.

- Nodes can be enlarged to cover as much of the object's area as possible. Such enlarged nodes are often used for moving the whole object.

- A cover may consist of the nodes that are not moved individually, but an attempt to move such a node results in moving of the whole object (the MoveNode() method only calls the Move() method). This makes an object moveable, but not resizable.

- Each of the nodes has its own parameters. It is easy to allow resizing along one direction but prohibit it along another; this means organizing a limited reconfiguration.

- It doesn't matter that some covers may represent the graphical objects and others controls or groups of controls. All covers are treated in the same way, thus allowing the user to change easily the inner view of any application.

- Covers can be visualized, though the best design makes moving / resizing of objects obvious without such visualization. Visualization of cover includes possible filling of the nodes' inner area and drawing of the nodes' perimeter.

## *Moving / resizing summary*

- To make any graphical object moveable / resizable, it must be derived from the `GraphicalObject` class and three methods must be written for such object: DefineCover(), Move(), and MoveNode().

- DefineCover() method defines the cover of an object as a set of nodes; each node has an individual number.

- Move() method describes the forward movement of an object as a whole. In reality it means the simple change of one or several primitives (points, rectangles).

- MoveNode() method describes the individual movement of the nodes; if the moving of a node results in the moving of a whole object, then the Move() method is called from inside the MoveNode() method. An individual movement of a node may cause the relocation of some or even all other nodes; in such cases the DefineCover() method is often called from inside the MoveNode() method.

- To organize the moving / resizing process, there must be an object of the `Mover` class.

    ```
    Mover mover;
    ```

- To prevent the accidental moving of elements out of view, the mover must be initialized with an additional parameter

    ```
    mover = new Mover (this);
    ```

- Three levels of moving the objects across the form's borders can be organized: moving outside is not allowed, moving allowed only across the right and lower borders, or moving allowed across all four borders.

- Mover has a queue of moveable objects and will supervise the whole moving / resizing process only for the objects that are included into this queue.

    ```
    mover .Add (…);
    mover .Insert (…);
    ```

- To make any control moveable / resizable, it's enough to include it into the mover's queue. Those three methods are not needed for controls.

- To make any control resizable, the appropriate values must be set to its `MinimumSize` and `MaximumSize` properties.

- Any combination of elements can organize a set of synchronously moving objects.

- For the complicated objects, consisting of the parts, which can be moved both synchronously and independently, and for the objects, for which the set of such parts can be changed, it is much better to develop an `IntoMover()` method, which is used instead of manual registering of all the moveable parts and which guarantees the correct registering of an object and all its parts regardless of a set of constituents.

- Moving and resizing are done with a mouse, and the whole process is organized via three standard mouse events: `MouseDown`, `MouseUp` and `MouseMove`.

- `MouseDown` starts moving / resizing of an object by grabbing one of the nodes in its cover. The only mandatory line of code in this method is

    ```
    mover .Catch (…);
    ```

- `MouseUp` ends moving / resizing by releasing any object that could be involved in the process.  The only mandatory line of code in this method is

    ```
    mover .Release ();
    ```

- `MouseMove` moves the whole object or a part of it.  There is one mandatory line of code in this method, but in order to see the movement, the `Paint` method must be called

    ```
    if (mover .Move (e .Location))
    {
        Invalidate ();
    }
    ```

- Mover can provide a lot of information about the object, which is currently moved or just released, and even about an object that is simply underneath the mouse cursor.  This data can be used to change the order of objects on the screen, to call the context menus, and so on.

- If covers must be shown, then one of the available drawing methods can be used, for example,

    ```
    mover .DrawCovers (grfx);
    ```

- If needed, several `Mover` objects can be used to organize the whole moving / resizing process.  Each mover deals only with the objects from its own queue.

## Conclusion

This is a brief (concentrated) version of my ideas about turning the screen objects into moveable / resizable.  The step from the objects, which behaviour is absolutely determined by the designer, to the objects, which can be moved and resized by the users, means a big change in design of applications.  For example, if the users get a chance to change a lot of (all!) parameters (positions, sizes, number of objects, their order,…), then there must be an absolutely reliable mechanism of saving and restoring all those objects and their parameters.  This article is only about the algorithm of turning objects into moveable / resizable, so I decided not to discuss this item here.  For the same reason, the saving / restoring is not included into the accompanying application, though it works in the **Form_NewComment.cs**, where the configuration is saved on the Registry and restored on the next opening of the form.  This mechanism is implemented in all the forms for demo application, which will come with the second article and will be described in that article.  The same mechanism works in nearly every form of the mentioned **Test_MoveGraphLibrary** application.

Moveable / resizable objects are the basis for design of the user-driven applications.  Such applications show improvement in comparison with the currently used fixed applications in nearly every area.  But there are areas, for example, the engineering / scientific or financial applications, in which the design on the basis of moveable elements means a step to another level of development and use.

This article is mostly about the algorithm of turning the screen objects into moveable / resizable.  It's a very important task by itself, but my work in this area showed me absolutely clear that the consequences of applying such algorithm to the design of applications are much more important than the algorithm itself.  It doesn't matter either you use this algorithm or another; in any way you'll come to the same understanding and the same conclusion: the moveable / resizable elements can't coexist with the fixed objects.  If you start using the moveable objects in applications, they will require, demand, and force you into redesign of everything around on the basis of exclusively moveable elements.  And as a result you'll come to the different kind of programs.

But this will happen, if you care about the users of your applications and think that they have to receive a chance to organize the applications in the way, which is the best personally for them.  Not as a crowd, but personally for each of them.  If you think that you know better than anyone of them, what they need, you can put aside all these new ideas and continue to work in the standard way of the fixed design and explain to your users that they have to be happy with you view once and for all.

As for myself, I see all the time how quickly the users of the user-driven applications got used to the new possibilities and expect this behaviour from all the new programs without questions.  The design of user-driven applications and the consequences of such design will be the theme of the second article.

Dr. Sergey Andreyev ( andreyev_sergey@yahoo.com )

December 2009 - January 2010

# Programs and documents

Several programs and documents are designed for better explanation of moveable and resizable graphics and its use. All files are available at www.SourceForge.net in the project **MoveableGraphics** (names are case sensitive there!). The most important files are renewed from time to time (usually every month); others can be older.

| | |
|---|---|
| **TheoryOfMoveableObjects.zip** | An application (the whole project with all the codes in C#) to accompany the current article. The article (current document) is also included into this ZIP file both in DOC and PDF formats. To run an application, only two files are needed: **TheoryOfMoveableObjects.exe** and **MoveGraphLibrary.dll.** |
| **Moveable_Resizable_Objects.doc** | The detailed description of the design and use of moveable / resizable objects. The explanation is based on the samples and codes from the **Test_MoveGraphLibrary** project. 96 pages. |
| **Test_MoveGraphLibrary.zip** | Contains the whole project from Visual Studio with a lot of samples and useful code. The source files are written in C#; all the samples in the **Moveable_Resizable_Objects.doc** are from this project. If you want to run this application, then only two files are needed: **Test_MoveGraphLibrary.exe** and **MoveGraphLibrary.dll.** The library is included into this ZIP file, but also is available as a stand alone file. |
| **MoveGraphLibrary.dll** | The library. |
| **MoveGraphLibrary_Classes.doc** | Description of the classes, included into the **MoveGraphLibrary.dll**. 115 pages. |
| **LiveCalculator.zip** | Calculator, which is included into the **Test_MoveGraphLibrary.exe** for explanation, but presented here as a separate program. To use this application, it must be accompanied by the DLL, which is also inside the zip file. |
| **MoveGraphLibrary_Graphics.doc** | Description of plotting, implemented in the **MoveGraphLibrary.dll**; description of all the tuning dialogs, which are also included into this library. |
| **TuneableGraphics.exe** | This application demonstrates the moveable / resizable objects from absolutely different areas. A year ago it was absolutely different application; now a lot of its forms are similar to the forms from the **Test_MoveGraphLibrary** application, but some are different. |
| **TuneableGraphics_Description.doc** | Description of the **TuneableGraphics** application. |